\documentclass[reprint,superscriptaddress,preprintnumbers,nofootinbib,nobibnotes,amsmath,amssymb,aps,prd,floatfix]{revtex4-2}

\usepackage{amsbsy,esint,mathrsfs,commath,amsfonts,mathtools,amsthm}
\usepackage[usenames,dvipsnames]{xcolor}
\usepackage{braket}
\usepackage{tensor}
\usepackage{physics}
\usepackage{graphicx}
\usepackage{xparse}
\usepackage{hyperref}
\usepackage[caption=false]{subfig}

\usepackage{tikz}

\usepackage[normalem]{ulem}

\hypersetup{
    colorlinks=true,
    linkcolor=Blue,
    filecolor=magenta,      
    urlcolor=Blue,
    citecolor=Blue,
    }

\newcounter{mnotecount}[section]

\renewcommand{\themnotecount}{\arabic{mnotecount}}
\newcommand{\mnote}[1]
{\protect{\stepcounter{mnotecount}}$^{\mbox{\footnotesize
$
\bullet$\themnotecount}}$ \marginpar{
\raggedright\tiny\it
$\!\!\!\!\!\!\,\bullet$\themnotecount: #1} }

\def\sa{\sigma^\alpha}
\def\sb{\sigma^\beta}

\def\SU{\mbox{SU}}
\def\SO{\mbox{SO}}
\def\U{\mbox{U}}

\def\nnu{\nu}

\newcommand{\boxbra}[1]{\left[ #1 \right|}
\newcommand{\boxket}[1]{\left| #1 \right]}

\NewDocumentCommand{\linknodo}{m o o}{%
    #1_{\IfValueT{#2}{\ensuremath{#2}}}^{\IfValueT{#3}{\ensuremath{#3}}}
  }

\begin{document}
 
\title{Area bounds and gauge fixing: alternative canonical variables for loop gravity}

 \date{\today}
  
  \author{I\~naki Garay} 
  \email{inaki.garay@ehu.eus}
  \affiliation{Department of Physics and EHU Quantum Center,
    University of the Basque Country UPV/EHU, Barrio Sarriena s/n, 48940, Leioa, Spain}  
  \author{Sergio Rodr\'iguez-Gonz\'alez}
  \email{sergio.rodriguez@ehu.eus}
  \affiliation{Department of Physics and EHU Quantum Center,
    University of the Basque Country UPV/EHU, Barrio Sarriena s/n, 48940, Leioa, Spain} 
  \author{Ra\"ul Vera} 
  \email{raul.vera@ehu.eus}
  \affiliation{Department of Physics and EHU Quantum Center,
    University of the Basque Country UPV/EHU, Barrio Sarriena s/n, 48940, Leioa, Spain}  

\begin{abstract}
 We use a canonical parametrization of twisted geometries describing the classical phase space of loop quantum gravity on a fixed graph, and establish its explicit correspondence with the associated frame bases and spinorial descriptions. Applied to the two-vertex model, this framework yields analytical bounds on the evolution of the total area, proving the existence of a non-vanishing lower bound at finite times. These findings, previously observed only numerically, suggest a bounce-like behavior and highlight the usefulness of these variables for the study of more general configurations. As a second result, the canonical variables are shown to simplify the gauge-fixing procedure, generalizing previous results restricted to two-vertex models with four links.
\end{abstract}

\maketitle

\tableofcontents

\section{Introduction}

Loop quantum gravity (LQG) provides a non-perturbative and background-independent framework for the quantization of gravity, where the classical phase space is described in terms of holonomy–flux variables defined on a graph \cite{Rovelli,Thiemann}. A particularly useful parametrization of this phase space is given by the twisted geometries framework, which offer a discrete description of space in terms of polyhedra associated with the nodes of the graph \cite{TwistedGeometries,TwistorsTG}. In this picture, the geometry is encoded in areas, normals to faces, and twist angles that characterize how adjacent polyhedra are glued together.

The spinorial formalism has proven to be a powerful tool to describe this structure, as it provides a convenient parametrization of the holonomy–flux algebra in terms of spinors \cite{returnSpinor,Tambornino1,Tambornino2,Tambornino3}. This approach has been especially fruitful in the study of dynamics, for instance in the context of the two-vertex model, where explicit Hamiltonians can be implemented and the evolution of the system can be analyzed in detail. It has been also shown that 
 this framework is well suited to explore symmetry-reduced sectors, including homogeneous, anisotropic, and inhomogeneous configurations \cite{returnSpinor,Rovelli2vertex,2vertexN,livinebenito,Eneko,Alvaro,garay_two-vertex_2025,FrameBasis,Assanioussi:2026cee}.

In this paper we use a canonical parametrization of twisted geometries formulated in terms of a set of variables that we denote as $\zeta$-variables (section \ref{section:newparam}). 
The motivation for introducing this parametrization is twofold. First, it allows us to extract analytical information about the dynamics that was previously only accessible through numerical analysis \cite{Eneko}. 
In particular, we show the existence of bounds for the total area under time evolution in the two-vertex model (section \ref{Sec:dynamics}). Second, the canonical variables are used to generalize previous parametrizations introduced to study the gauge invariant phase space of two-vertex graphs with only four links  \cite{garay_two-vertex_2025}  to a general graph.

\section{Alternative parametrization of twisted geometries}\label{section:newparam}

In this section we review the spinorial formalism and the twisted geometries framework for LQG in order to later introduce a canonical parametrization of them that will provide a useful geometrical picture of the classical phase space of LQG.

\subsection{Spinorial formalism and twisted geometries in LQG}

The classical degrees of freedom of LQG on a fixed graph are encoded in the holonomy-flux variables, defined as pairs $(g,X)\in \SU(2)\times \mathfrak{su}(2)$ that are located at the links joining the nodes of the graph \cite{Rovelli,Thiemann}. The Poisson brackets are given by
\begin{align}
\hspace*{-3mm}\left\{g_{AB},g_{CD}\right\} &= 0,  \label{HolonomyFluxAlgebra1}\\
\left\{X^I ,X^J \right\} &= {\epsilon}{^{IJK}}X^{K},\quad\!
    \left\{\vec{X},g_{AB}\right\} = -i\frac{1}{2}(g\vec{\sigma})_{AB},
    \label{HolonomyFluxAlgebra2}
\end{align}
where $A,B,\ldots = 0,1$ denote components in the defining representation of $\SU(2)$; $\vec{X}= \frac{1}{2}\Tr\{X \vec{\sigma}\}$, so that $I,J,\ldots = 1,2,3$ are indices in $\mathbb{R}^{3}$; and $\sigma^1,\sigma^2,\sigma^3$ are the Pauli matrices.

The spinorial formalism of LQG
\cite{returnSpinor,Tambornino1,Tambornino2,Tambornino3} offers a
convenient description of the holonomy-flux
algebra in which the classical degrees of freedom are encoded in
spinors $\ket{z}\in \mathbb{C}^{2}$. In particular, since the graph is
oriented, its links are directed from a source node to a target node,
each of which is assigned a spinor, $\ket{z^{s}}$ and $\ket{z^{t}}$,
respectively. These and their conjugates are expressed in terms of
their components $z^{A} =z^{0},z^{1}\in \mathbb{C}$ by
\begin{equation}
    \ket{z} = \begin{pmatrix}
        z^{0} \\ z^{1}
    \end{pmatrix}, \hspace{5mm} \bra{z} = \begin{pmatrix}
        \overline{z^{0}} & \overline{z^{1}}
    \end{pmatrix}.
\end{equation}
Also, for each spinor $\ket{z}$, a dual spinor $\boxket{z}$ satisfying $\boxbra{z}z\rangle = 0$ and $\boxbra{z}z] = \braket{z}$ is defined according to
\begin{equation}
    \boxket{z} = \begin{pmatrix}
        -\overline{z^{1}} \\ \overline{z^{0}}
    \end{pmatrix}, \hspace{5mm} \boxbra{z} = \begin{pmatrix}
        -{z}^{1} & {z}^{0}
    \end{pmatrix}.
\end{equation}
These spinors parametrize the holonomy-flux pair through the equations
\begin{equation}
    \Vec{X}^{s,t}\! :=\! \frac{1}{2}\! \bra{z^{s,t}} \Vec{\sigma} \ket{z^{s,t}},\qquad\!\!  g := \frac{|z^t] \bra{z^s} - \ket{z^t} [z^s|}{\sqrt{\braket{z^t} \braket{z^s}}},\label{DEF vectores normales}
\end{equation}
when the flux is identified as $\vec{X} = \vec{X}^{s}$ and $\braket{z^{s}},\braket{z^{t}} \neq 0$. It is direct to check that $X^{s,t} := \vec{X}^{s,t}\cdot \vec{\sigma}$ are related, by construction, through the equation
\begin{equation}
    \frac{X^{t}}{|\vec{X}^{t}|} =  -g \frac{X^{s}}{|\vec{X}^{s}|} g^{-1}=:-\text{Ad}_{g}\left(\frac{X^{s}}{|\vec{X}^{s}|}\right) , \label{parallel_transport}
\end{equation}
so that $\vec{X}^{t}$ can be interpreted as the result of  performing a parallel transport of $\vec{X}^{s}$ along the link by means of $g$.

Observe that this parametrization introduces two extra real parameters describing the classical degrees of freedom in a link, which enlarges the gauge freedom of the theory. In fact, the $\U(1)$ transformations
\begin{equation}
    \left(\ket{z^{s}},\ket{z^{t}}\right) \to \left(e^{i\phi}\ket{z^{s}},e^{-i\phi}\ket{z^{t}}\right), \hspace{2.5mm} \phi \in \mathbb{R}, \label{U1transformations}
\end{equation}
leave equations \eqref{DEF vectores normales} unchanged. Let us
define the matching constraint as
\begin{equation}
    \mathcal{C} := \braket{z^{s}} - \braket{z^{t}},\label{matching}
\end{equation}
and equip $\mathbb{C}^2\times\mathbb{C}^2$ with the Poisson brackets
\begin{equation}
    \left\{z^{sA},\overline{{z}^{sB}}\right\} = \left\{z^{tA},\overline{{z}^{tB}}\right\} = -i\delta^{AB},
    \label{poissonbrackets}
\end{equation}
while the rest of them vanish. It can be checked that $\mathcal{C}$ is the infinitesimal generator of the $\U(1)$ transformations \eqref{U1transformations}.
Imposing $\mathcal{C} = 0$
is the necessary and sufficient condition to recover the holonomy-flux algebra \eqref{HolonomyFluxAlgebra1}-\eqref{HolonomyFluxAlgebra2} from the definitions \eqref{DEF vectores normales} and \eqref{poissonbrackets},
and, fixing the associated $\U(1)$ gauge, 
reduces the eight spinorial degrees of freedom to the six encoded in $(g,X)$.

To extend this phase space 
to the whole graph 
we impose the Poisson brackets between spinors at different links to vanish, while
the Gauss constraint at each node $\nu$
takes the form of the closure constraints,
\begin{equation}
    \vec{\mathcal{X}}^{\nu} := \sum_{i \in \nu}\vec{X}^\nu_{i} = 0,\label{DEF closure constraint}
\end{equation}
where 
the sum is extended to all the links, labelled by $i,j,\ldots$, joining $\nu$.
The closure constraints at a node $\nu$ induce the $\SU(2)$ transformations $\ket{z^{\nu}} \to h^\nnu \ket{z^{\nu}}$, with $h^\nnu\in  \SU(2)$, and both the matching and closure constraints constitute a system of Poisson-commuting constraints.

The matching and closure constraints, together with the relation \eqref{parallel_transport}, give rise to an interpretation of the spinorial formalism on a graph as a discretization of space in terms of polyhedra. 
Indeed, a set of nonzero vectors $\vec{X}^{\nu}_i$ spanning $\mathbb{R}^3$ and satisfying \eqref{DEF closure constraint} are in one to one correspondence with the faces of a unique (in shape) convex polyhedron:
$\vec{X}^{\nu}_i$ are normal to the faces, pointing outwards, and
the  norms  $|\vec{X}^{\nu}_i|$ equal their corresponding areas \cite{Alexandrov2005}.
This establishes a correspondence between the nodes of the graph and convex polyhedra, which allows to interpret the nodes as the classical description of the quanta of volume in LQG. 
The matching constraints can then be interpreted in these terms as the equality of the areas of two adjoining faces of two linked polyhedra. However, the matching constraints say nothing about the shape of these two faces, 
so that polyhedra are glued together in a \textit{twisted} way, thus generalizing the Regge geometries. 

\begin{figure}
    \centering
    \includegraphics[width=0.45\textwidth]{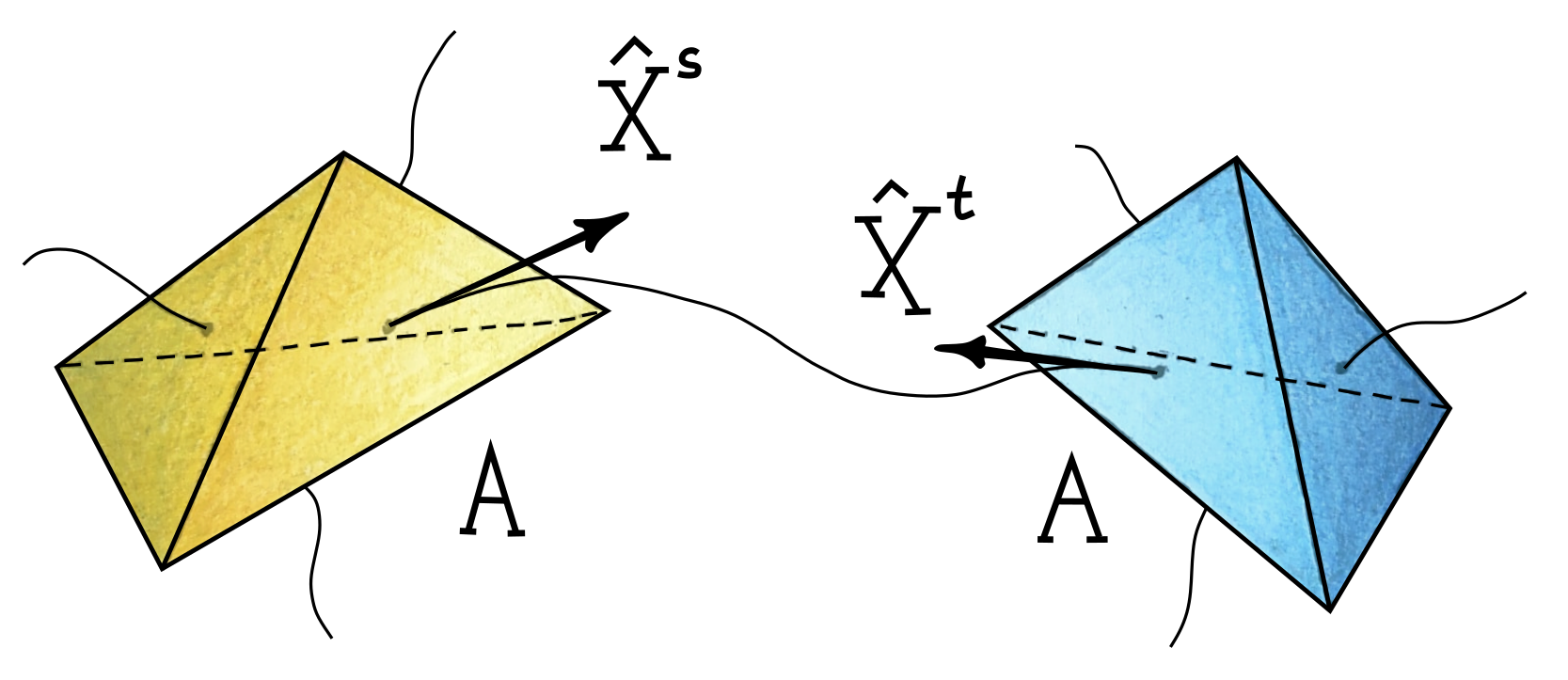}
    \centering
    \caption{The twisted geometries variables $A, \hat{X}^{s},\hat{X}^{t}$ that determine the area and orientation of two contiguous faces of two adjoining polyhedra.}
    \label{figure_twisted_variables}
\end{figure}

\begin{figure}
    \centering
    \includegraphics[width=0.5\textwidth]{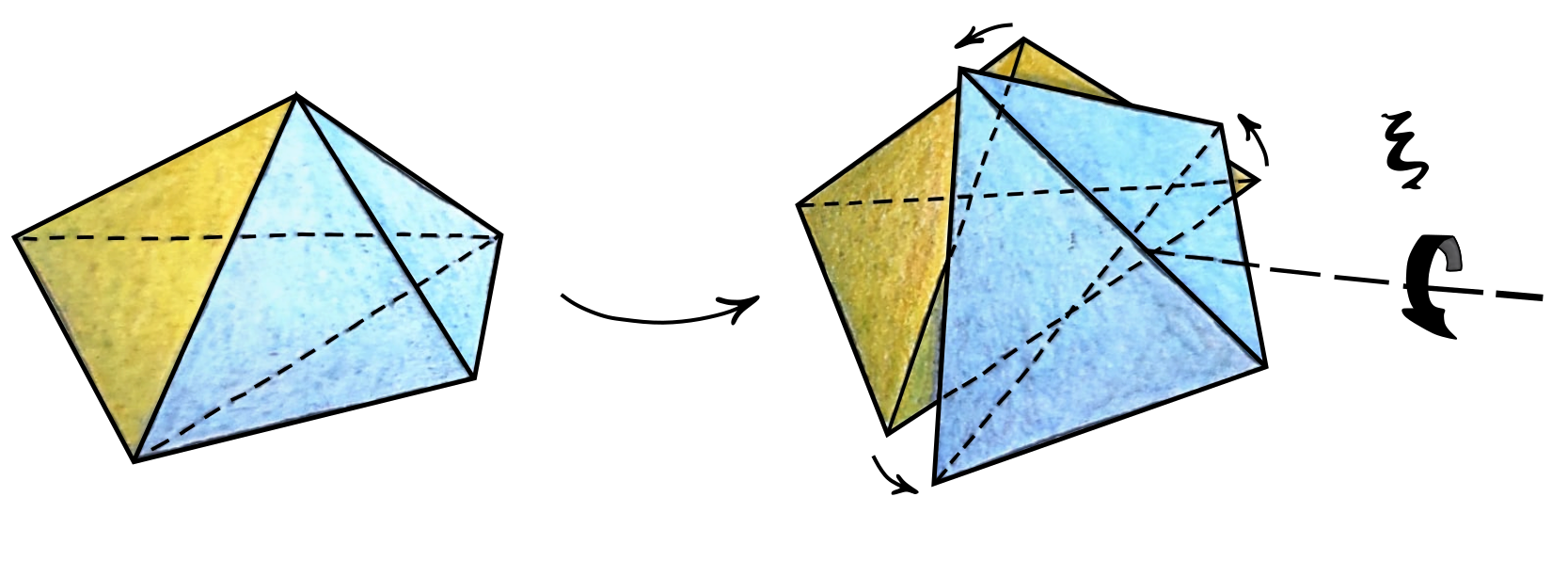}
    \centering
    \caption{The twist angle $\xi$ associated with two contiguous faces of two adjoining polyhedra.}
    \label{figure_twisting}
\end{figure}

This kind of discretization of space, known 
as twisted geometries, can be described at each link of the graph by the quantities $(A_i,\hat{X}_{i}^{s},\hat{X}_{i}^{t},\xi_i)$, where $A_i$ represents the area of the faces of the polyhedra joining at the link, $\hat{X}_{i}^{s},\hat{X}_{i}^{t}$ are the unit vectors in the direction of the flux vectors at the source and target nodes, and $\xi_i\in[-\pi,\pi]$ is the so-called twist angle (see figure \ref{figure_twisted_variables}) \cite{TwistedGeometries,TwistorsTG}. 
The first three variables allow us to reconstruct $\vec{X}^{s}_{i}, \vec{X}_{i}^{t}$ and hence all the polyhedra attached to the nodes of the graph, given that the closure constraint is satisfied. The twist angle is the extra information specifying how polyhedra are \textit{glued} to each other (see figure \ref{figure_twisting}) \cite{TwistedGeometries}. 

This polyhedral decomposition can be also described purely in terms of three-dimensional vectors and $\SO(3)$ rotations by the frame bases, which are sets of orthogonal vectors located at the faces of the polyhedra on the graph \cite{FreidelCS,FrameBasis}. Denoted by $\{\Vec{X}_{i},\Vec{G}_{i},\Vec{F}_{i}\}$, they are defined in terms of the spinors by
    \begin{align}
        \Vec{X}_{i} &= \frac{1}{2}\bra{z_i}\Vec{\sigma}\ket{z_i},\label{def normales} \\ \Vec{G}_{i} &= \frac{1}{2}\Re{\boxbra{z_i}\Vec{\sigma}\ket{z_i}}, \\ \Vec{F}_{i} &= \frac{1}{2}\Im{\boxbra{z_i}\Vec{\sigma}\ket{z_i}},\label{def efes}
    \end{align}
where the index $i$ labels the face of the polyhedron. It is direct to show that the frame basis determines its associated spinor, up to its global sign.  In the next section we will introduce a parametrization of the spinorial formalism of LQG that connects all these different representations in a straightforward manner and gives rise to canonical Poisson bracket relations.

\subsection{A canonical parametrization of the frame basis}

\begin{figure}
    \centering
    \includegraphics[width=0.4\textwidth]{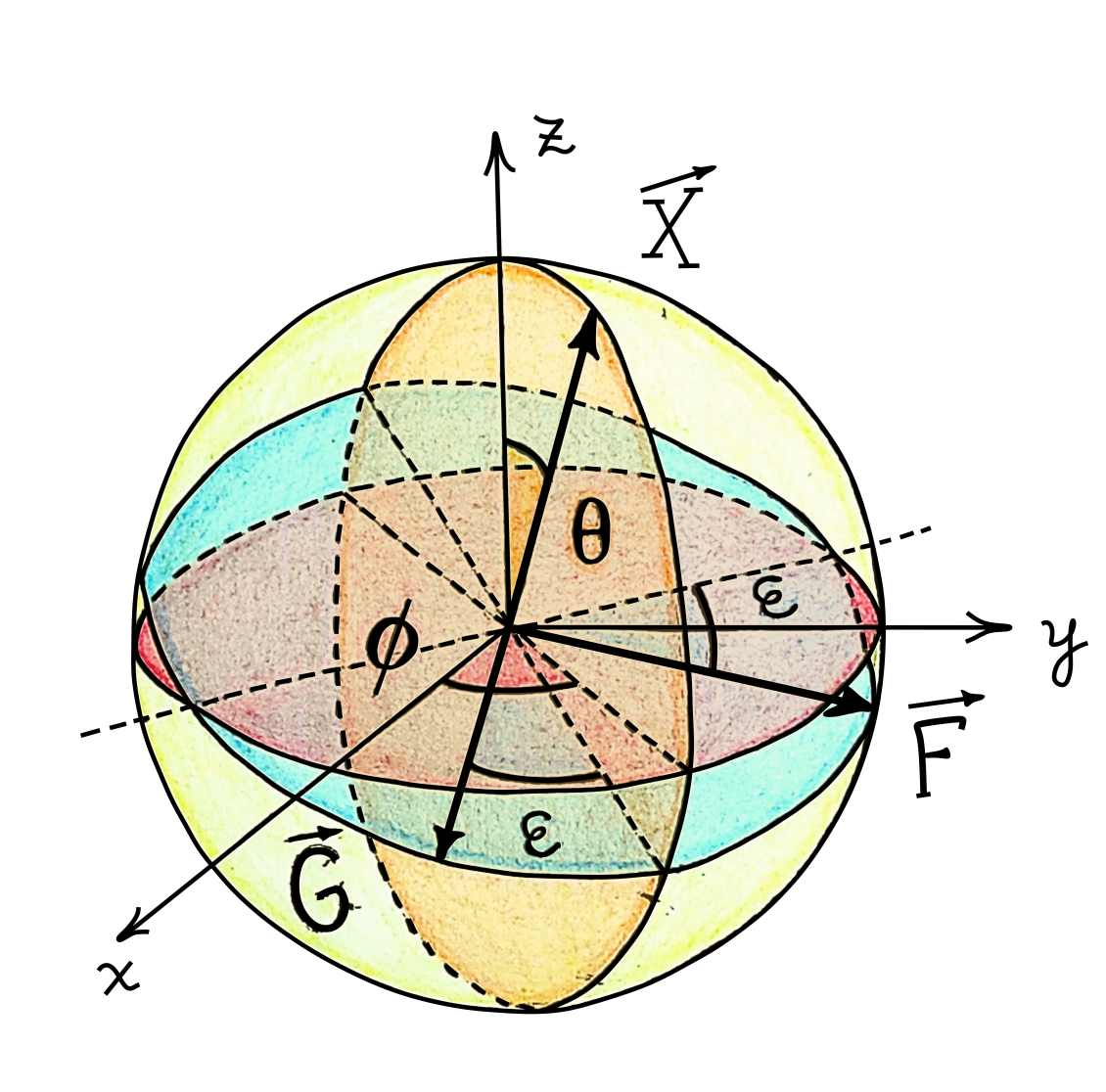}
    \centering
    \caption{Euler angles $\phi,\theta,\varepsilon$ determining the direction of the frame basis vectors.}
    \label{figura_euler_angles}
\end{figure}

The equations \eqref{def normales}-\eqref{def efes} constitute the standard representation of a triad in terms of spinors, which is also known to be related to its parametrization in terms of its Euler angles \cite{tisza2005applied}. Such relation is given by the equations
\begin{gather}
    z^0 = \sqrt{2R}\cos{\left(\frac{\theta}{2}\right)}e^{\frac{i}{2}(\varepsilon-\phi)},\label{z0_theta}\\ z^1 = \sqrt{2R}\sin{\left(\frac{\theta}{2}\right)}e^{\frac{i}{2}(\varepsilon+\phi)}\label{z1_theta}.
\end{gather}
Indeed, it is direct to check that introducing these expressions into the equation \eqref{def normales} yields the parametrization
\begin{align}
    X_{1} &= R\sin{\theta}\cos{\phi},\label{componentes X}\\
    X_{2} &= R\sin{\theta}\sin{\phi},\\
    X_{3} &= R\cos{\theta},\label{componentes X 3}
\end{align}
so that $R$ is the modulus of $\vec{X}$ (equal to the area of the
face), $\theta$ and $\phi$ are the polar and azimuthal angles in the
spherical representation of $\vec{X}$, and, hence, $\varepsilon$ is
the angle that contains the information about the position of
$\vec{G}$ and $\vec{F}$ in the plane normal to $\vec{X}$ (see figure
\ref{figura_euler_angles}). The fact that the global sign of a spinor
cannot be recovered from the associated frame basis is a consequence
of the fact that performing transformations of the kind $\varepsilon
\to \varepsilon + 2\pi n$ or $\phi \to \phi + 2\pi m$, with $n,m \in
\mathbb{Z}$, modifies the sign of the right hand side of the equations
\eqref{z0_theta}-\eqref{z1_theta}. As a result, the variables
$R,\theta,\phi\mod{2\pi}, \varepsilon \mod{2\pi}$ determine the frame basis, but not the
associated spinor.

As we are going to see next, it is useful to
change the variable $\theta$ by
\begin{equation}
    \zeta = R\cos{\theta} \hspace{2.5mm} \longleftrightarrow \hspace{2.5mm} \theta = \arccos{(\zeta/R)},\label{zeta}
\end{equation}
where we fix $\arccos$ to return a value in $(-\pi,\pi]$. In terms of the new variable $\zeta$,
\eqref{z0_theta}-\eqref{z1_theta} become
\begin{equation}
    z^0 = \sqrt{R + \zeta}e^{\frac{i}{2}(\varepsilon-\phi)},\quad z^1 = \sqrt{R - \zeta}e^{\frac{i}{2}(\varepsilon+\phi)}\label{z1},
\end{equation}
which reproduce previously used expressions for the spinor components \cite{Assanioussi:2026cee,returnSpinor,Livine:2013tsa,Dupuis:2011fz}. Therefore, when $z^0\neq0$ and $z^{1} \neq 0$, the
set of variables $\{R,\varepsilon,\phi,\zeta\}$ is
given in terms of the spinor components by
\begin{align}
R &= \frac{1}{2}\left(|z^0|^2 + |z^1|^2\right),
&\hspace{2.5mm}
    \phi &= \arg{z^1} - \arg{z^0},\label{NewVariables_2}
    \\
    \varepsilon &= \arg{z^0} + \arg{z^1}, &\hspace{2.5mm}
    \zeta &= \frac{1}{2}\left(|z^0|^2 - |z^1|^2\right).\label{inverse epsilon phi}
\end{align}
On the other hand, if $z^0=0$ or $z^{1} = 0$, the complex function
$\arg$ is undefined and thus equations \eqref{NewVariables_2}-\eqref{inverse epsilon
phi} are no longer valid for $\varepsilon$ and $\phi$. This ambiguity
comes from the fact that the transformations $(\varepsilon,\phi) \to
(\varepsilon + \lambda,\phi-\lambda)$ when $z^{0}=0$ and
$(\varepsilon,\phi) \to (\varepsilon + \lambda,\phi+\lambda)$ when
$z^{1}=0$, for any $\lambda \in \mathbb{R}$, do not modify the right
hand side of equations \eqref{z1}. Therefore, in order to
uniquely determine $\varepsilon$ and $\phi$ from $\ket{z}$ in these
singular cases, one has to fix a choice. By convenience, we extend the
inverse transformation \eqref{inverse epsilon phi} to these two
degenerate cases by taking $\varepsilon = \phi$ when $z^{0}=0$ and
$\varepsilon = -\phi$ when $z^{1}=0$ (as explained in the previous
section, we are not concerned with the case $z^{0} = z^{1} = 0$).

One of the virtues of working with this parametrization is that
when $z^{0},z^{1} \neq 0$ the set of variables
$\{R,\varepsilon,\phi,\zeta\}$ satisfies the Poisson brackets
\begin{equation} \left\{R,\varepsilon\right\} =
\left\{\phi,\zeta\right\} = 1,\label{Poisson_brackets_NewVariables}
\end{equation} so that they are canonical. Moreover, even when the
transformation from $\ket{z}$ to $\{R,\varepsilon,\phi,\zeta\}$
is not continuous when $z^0 = 0$ or $z^1 = 0$, the limit of the
Poisson brackets \eqref{Poisson_brackets_NewVariables} when
approaching these points is well defined and equal to $1$, so that we
can naturally extend the equality
\eqref{Poisson_brackets_NewVariables} to the whole domain of
definition of the variables $\{R,\varepsilon,\phi,\zeta\}$.  We
will refer to the set $\{R,\varepsilon,\phi,\zeta\}$ as the
$\zeta$-variables in the remainder.

\subsection{$\zeta$-variables and twisted geometries}
\label{sec:variables}
We have already seen
that the $\zeta$-variables completely determine the frame basis attached
to a polyhedron face. Since the frame bases attached to two adjoining
faces of two contiguous polyhedra completely determine the associated
twisted geometries variables \cite{FrameBasis}, we can also find an
explicit relation between the twisted geometries variables and
the $\zeta$-variables.

With this purpose, let us focus on two adjoining faces of two contiguous polyhedra in the
twisted geometries formalism. These two faces are associated to an oriented link of the underlying graph, where the two spinors $\ket{z^s}$ and $\ket{z^t}$ are placed.
There, the six independent twisted geometries variables
\begin{equation}
    \left\{A,\linknodo{\hat{X}}[][s],\linknodo{\hat{X}}[][t],\xi\right\}\label{twisted variables}
\end{equation}
introduced above already take into account the matching constraint \eqref{matching} and the associated gauge fixing. A priori, describing those faces with the $\zeta$-variables would imply working with the eight independent canonical variables
\begin{equation}
    \left\{\linknodo{R}[][s],\linknodo{\varepsilon}[][s],\linknodo{R}[][t],\linknodo{\varepsilon}[][t],\linknodo{\phi}[][s],\linknodo{\zeta}[][s],\linknodo{\phi}[][t],\linknodo{\zeta}[][t]\right\}.\label{variables a priori}
\end{equation}
However, as mentioned earlier, this set reduces to six variables after imposing the matching constraint \eqref{matching} and fixing the associated $\U(1)$ gauge freedom.
The matching constraint \eqref{matching} on the link takes the simple form
\begin{equation}
    \mathcal{C} = 2\left(\linknodo{R}[][s] - \linknodo{R}[][t]\right),\label{matching RsRt}
\end{equation}
and generates, through Poisson brackets, gauge transformations of the kind $(\linknodo{\varepsilon}[][s],\linknodo{\varepsilon}[][t]) \to (\linknodo{\varepsilon}[][s] - \lambda,\linknodo{\varepsilon}[][t] + \lambda)$, with $\lambda \in \mathbb{R}$, that trivially imply the $\U(1)$ transformations in \eqref{U1transformations}. We can then define the functions
\begin{equation}
    A := \frac{1}{2}(\linknodo{R}[][s]+\linknodo{R}[][t]), \hspace{5mm} \Phi := \linknodo{\varepsilon}[][s] + \linknodo{\varepsilon}[][t],\label{A y Phi}
\end{equation}
which are $\U(1)$ gauge invariant, and reduce the number of variables on the
link to the six independent canonical quantities 
\begin{equation}
\left\{A,\Phi,\linknodo{\phi}[][s],\linknodo{\zeta}[][s],\linknodo{\phi}[][t],\linknodo{\zeta}[][t]\right\}.\label{variables matching}
\end{equation}
We shall call this reduced set of six variables the reduced $\zeta$-variables.

The relation between the variables \eqref{variables matching} and the twisted geometries variables \eqref{twisted variables}
is the following.
The unit normal vectors $\hat{X}^s$ and $\hat{X}^t$ are recovered from $\phi^s,\zeta^s,\phi^t,\zeta^t$
using equations \eqref{componentes X}-\eqref{componentes X 3} and \eqref{zeta}
with $R=1$, the function $A$
coincides with the area $A$ in the twisted geometries,
and the twist angle $\xi$
is given by any one of the four possibilities encoded in the object defined by
\begin{equation}
    \linknodo{\xi}[][\sigma\tilde\sigma] := \Phi - \sigma \linknodo{\phi}[][s] -\tilde\sigma\linknodo{\phi}[][t],\label{def:xi}
\end{equation}
where $\sigma,\tilde\sigma \in \left\{-1,1\right\}$. 
Any of the four possibilities is valid, in principle, and different options
  have been used in the literature. For instance, the two definitions taken in
\cite{TwistorsTG},
later used in \cite{Alvaro}, correspond to the two choices that satisfy $\sigma = \tilde\sigma$. More concretely, $\xi_0$ and $\xi_1$ in \cite{TwistorsTG} correspond to
  $\xi_0=-\xi^{(-1)(-1)}$ and $\xi_1=-\xi^{(1)(1)}$.
  
\section{Dynamics of the two-vertex model in terms of the
  reduced $\zeta$-variables}
  \label{Sec:dynamics}
  
The reduced $\zeta$-variables, previously introduced  in section \ref{section:newparam}, allow us to easily extract geometric information about the evolution of the polyhedra associated to the nodes in a general graph. In the case of the simplified model given by the truncation of LQG known as the two-vertex model 
\cite{returnSpinor,Rovelli2vertex,2vertexN,livinebenito,Eneko,Alvaro,garay_two-vertex_2025,FrameBasis}, the reduced $\zeta$-variables provide a natural geometric way of dealing with the dynamics of the model and with the gauge fixing procedure. In fact, they generalize the variables introduced in \cite{garay_two-vertex_2025} for the specific case of four links.

In this section we use the $\zeta$-variables in order to further study the dynamics of the two-vertex model in the general case (arbitrary number of links and no symmetry reductions applied). The equations of motion will be explicitly obtained and  interesting results about the bounds of the area will be extracted.

\subsection{Two vertex model revisited}

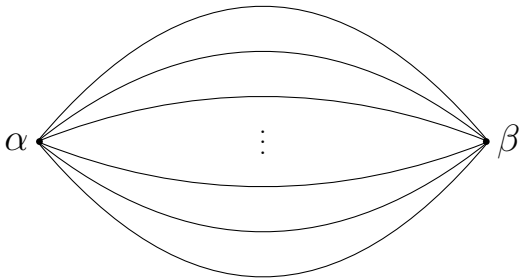
\begin{figure}[h]
\begin{tikzpicture}

\def\step{0.8}

\coordinate (A) at (0,0);
\coordinate (B) at (6,0);

\node[left] at (A) {{\Large $\alpha$}};
\node[right] at (B) {{\Large $\beta$}};

\draw (A) .. controls (2,3*\step) and (4,3*\step) .. (B);
\draw (A) .. controls (2,2*\step) and (4,2*\step) .. (B);
\draw (A) .. controls (2,\step) and (4,\step) .. (B);

\draw (A) .. controls (2,-3*\step) and (4,-3*\step) .. (B);
\draw (A) .. controls (2,-2*\step) and (4,-2*\step) .. (B);
\draw (A) .. controls (2,-\step) and (4,-\step) .. (B);

\node at (3,0.1) {$\vdots$};

\fill (0.03,0) circle (0.45mm);
\fill (5.97,0) circle (0.45mm);

\end{tikzpicture}
    \caption{The graph of the two-vertex model.}
    \label{figure_grafo}
  \end{figure}
The two-vertex model consists of a graph with two nodes $\nu$, denoted as
   $\alpha$ and $\beta$, connected by $N$ links, labelled by $i,j=1,\ldots,N$ (see figure \ref{figure_grafo}).
  {Its study has proven to be very useful in order to implement explicit dynamics,
  both quantum mechanical and classical, 
  and search through cosmological settings within the full theory \cite{Rovelli2vertex,livinebenito,returnSpinor,2vertexN,Eneko,FrameBasis,Alvaro,garay_two-vertex_2025}.
  
In order to implement the classical dynamics, different ansatz for a Hamiltonian constraint have been proposed in the model. In this paper, we  restrict ourselves to the Hamiltonian used in \cite{Alvaro,garay_two-vertex_2025,FrameBasis}  which is given in terms of the $\SU(2)$ invariant quantities (defined at each vertex)
  \begin{equation}
    E_{ij}:=\braket{z_i}{z_j},\quad F_{ij}:=[z_i\ket{z_j},\label{def:EF}
    \end{equation}
by
\begin{equation}
    H = \mathcal{N} \sum_{i,j = 1}^{N}\left(\lambda \linknodo{E}[ij][\alpha]\linknodo{E}[ij][\beta] + \Re{\gamma \linknodo{F}[ij][\alpha]\linknodo{F}[ij][\beta]}\right),\label{hamiltoniano}
\end{equation}
where $\mathcal{N}$ is the lapse function, that depends only on time, and $\lambda \in \mathbb{R}$ and $\gamma \in \mathbb{C}$ are coupling constants.

To compute the time evolution of any function of the spinorial or twisted geometries variables 
with this Hamiltonian via Poisson brakets let us 
first rewrite the quantities \eqref{def:EF} in terms of the $\zeta$-variables, partially reduced by using the matching constraint,
  so that $R^\alpha_i=R^\beta_i=A_i$.
At the node $\alpha$ we have
\begin{align}
    &\linknodo{E}[ij][\alpha] = \sum_{\sa = \pm1} \linknodo{x}[ij][\alpha]\, e^{\frac{i}{2}(\linknodo{\varepsilon}[j][\alpha] - \linknodo{\varepsilon}[i][\alpha])} e^{\frac{i}{2}\sa(\linknodo{\phi}[i][\alpha] - \linknodo{\phi}[j][\alpha])}\label{Eij nuevas variables},\\
     &\linknodo{F}[ij][\alpha] = \sum_{\sa = \pm1} \sa \linknodo{y}[ij][\alpha]\, e^{\frac{i}{2}(\linknodo{\varepsilon}[j][\alpha] + \linknodo{\varepsilon}[i][\alpha])} e^{\frac{i}{2}\sa(\linknodo{\phi}[j][\alpha] - \linknodo{\phi}[i][\alpha])}\label{Fij nuevas variables},
\end{align}
where we have introduced the notation
\begin{gather}
    \linknodo{x}[ij][\alpha]: = \sqrt{(A_i + \sa \linknodo{\zeta}[i][\alpha])(A_j + \sa \linknodo{\zeta}[j][\alpha])},\\
    \linknodo{y}[ij][\alpha]: = \sqrt{(A_i + \sa \linknodo{\zeta}[i][\alpha])(A_j - \sa \linknodo{\zeta}[j][\alpha])}.
\end{gather}
For the node $\beta$ the expressions are analogous.
Observe first that for any of the two nodes $\nu=\alpha,\beta$ we have
  $x^\nu_{ij}=x^\nu_{ji}$, and
  that since $|\zeta^\nu_i|\leq A_i$ by construction,
  see \eqref{zeta},
  then the square roots are well defined and furthermore
  \begin{align}\label{eq:bounds_xy}
    0\leq  y^\nu_{ij}\leq 2\sqrt{A_i A_j},\quad
        0\leq  x^\nu_{ij} \leq 2\sqrt{A_i A_j}
  \end{align}
  for any value of $\sigma^\nu$.
  Note finally that
  $\linknodo{x}[ii][\nu]= (A_i + \sigma^\nu \linknodo{\zeta}[i][\nu])$.

Introducing \eqref{Eij nuevas variables}-\eqref{Fij nuevas variables} in the Hamiltonian \eqref{hamiltoniano} we obtain
\begin{align*}
  &H = \mathcal{N}\sum_{i,j}\sum_{\sa,\sb}\Bigg\{\lambda \linknodo{x}[ij][\alpha]\linknodo{x}[ij][\beta]\cos{\left(\frac{\linknodo{\xi}[j][\sa\sb] - \linknodo{\xi}[i][\sa\sb]}{2}\right)} \\
  &+ |\gamma|\sa\sb \linknodo{y}[ij][\alpha]\linknodo{y}[ij][\beta]\cos{\left(\frac{\linknodo{\xi}[i][\sa\sb] + \linknodo{\xi}[j][(-\sa)(-\sb)]}{2} + \arg{\gamma}\right)}\Bigg\},
\end{align*}
where the sums are evaluated over $i,j \in\left\{1,\ldots,N\right\}$
and $\sa,\sb \in \left\{-1,1\right\}$, and
$\linknodo{\xi}[i][\sa\sb]$ are defined as in \eqref{def:xi} for each
link $i$.

Writing the Hamiltonian in terms of the $\zeta$-variables
has allowed us to obtain an expression involving only real valued $\U(1)$ gauge invariant quantities at each link
$\left\{A,\Phi,\phi^{\alpha},\zeta^{\alpha},\phi^{\beta},\zeta^{\beta}\right\}$,
that is, the set of reduced $\zeta$-variables.
Observe how $\phi^\alpha $, $\phi^\beta$ and $\Phi$ are encoded
  in the different twists $\xi$.
This shows, in particular,
the explicit dependence on the associated twist angles, which reflects
their role on the dynamics. Interestingly, the variables
$A,\zeta^{\alpha},\zeta^{\beta}$ and their conjugated angular
variables $\Phi,\phi^{\alpha},\phi^{\beta}$ get factored apart,
the first set encoded in $x^\nu_{ij}$ and $y^\nu_{ij}$
  and the second set encoded in the twists that appear
  as arguments of trigonometric functions, which greatly simplifies
the computation of the Poisson brackets needed to study evolution.

\subsection{Equations of motion in terms of the $\zeta$-variables}

The equations of motion computed in terms of the reduced $\zeta$-variables yield
\begin{widetext}
\begin{flalign}
  \frac{d A_k}{dt}
  =& \left\{A_k,H\right\} 
    =\mathcal{N}\!\!\!\sum_{i, \sa ,\sb}\!\!\left[
    \lambda x_{ik}^\alpha x_{ik}^\beta \sin(\frac{\linknodo{\xi}[i][\sa\sb] - \linknodo{\xi}[k][\sa \sb]}{2})\!-|\gamma|
\sa\sb
 y_{ki}^\alpha y_{ki}^\beta\sin(\frac{\linknodo{\xi}[k][\sa\sb] + \linknodo{\xi}[i][(-\sa)(-\sb)]}{2}+\text{arg}\gamma)    
    \right]\!,\label{eom:Ak}
\end{flalign}
  \begin{flalign*}
    \frac{d\linknodo{\Phi}[k]}{dt}
    =& \left\{\linknodo{\Phi}[k],H\right\}\nonumber\\
    =&-\mathcal{N}\lambda\sum_{i,\sa,\sb}x^\alpha_{ik}\frac{x^\beta_{ii}}{x^\beta_{ik}}
           \cos(\frac{\linknodo{\xi}[i][\sa\sb] - \linknodo{\xi}[k][\sa\sb]}{2})\nonumber&\\
     &-\mathcal{N}\frac{|\gamma|}{2}\!\sum_{i,\sa,\sb}
    \! \sa\sb
    \! \left[
       \frac{y^\alpha_{ki}}{ x^\alpha_{kk}}y^\beta_{ki}
          \cos(\frac{\linknodo{\xi}[k][\sa\sb]+\linknodo{\xi}[i][(-\sa)(-\sb)]}{2}+\text{arg}\gamma)
  +y^\alpha_{ik}\frac{x^\beta_{ii}}{y^\beta_{ik}}
       \cos(\frac{\linknodo{\xi}[i][\sa\sb] + \linknodo{\xi}[k][(-\sa)(-\sb)]}{2}+\text{arg}\gamma)\right]\nonumber\\&+\left(\alpha\longleftrightarrow\beta\right),
  \end{flalign*}
  \begin{flalign*}
    \frac{d\linknodo{\phi}[k][\alpha]}{dt}
    = &\left\{\linknodo{\phi}[k][\alpha], H\right\}\nonumber &\\
    =&\mathcal{N}\lambda \sum_{i, \sa ,\sb} \sa\frac{x^\alpha_{ii}}{x^\alpha_{ik}}x^\beta_{ik} 
       \cos(\frac{\linknodo{\xi}[i][\sa\sb] - \linknodo{\xi}[k][\sa\sb]}{2})\nonumber\\
&+\mathcal{N}\frac{|\gamma| }{2}\sum_{i, \sa ,\sb} 
\sb
  \left[ 
  \frac{y^\alpha_{ki}}{x^\alpha_{kk}} 
     y^\beta_{ki}
    \cos(\frac{\linknodo{\xi}[k][\sa\sb] + \linknodo{\xi}[i][(-\sa)(-\sb)]}{2}+\text{arg}\gamma)
   -\frac{ x^\alpha_{ii}}{y^\alpha_{ik} } y^\beta_{ik}
     \cos(\frac{\linknodo{\xi}[i][\sa\sb] + \linknodo{\xi}[k][(-\sa)(-\sb)]}{2}+\text{arg}\gamma)
    \right],
\end{flalign*}
\begin{flalign*}
  \frac{d\linknodo{\zeta}[k][\alpha]}{dt}
  =& \left\{\linknodo{\zeta}[k][\alpha],H\right\}
  =\mathcal{N}\sum_{i, \sa ,\sb}\left[
    \lambda \sa  x_{ik}^\alpha x_{ik}^\beta \sin(\frac{\linknodo{\xi}[i][\sa\sb] - \linknodo{\xi}[k][\sa \sb]}{2}) -|\gamma|
     \sb
      \,y_{ki}^\alpha y_{ki}^\beta\sin(\frac{\linknodo{\xi}[k][\sa\sb] + \linknodo{\xi}[i][(-\sa)(-\sb)]}{2}+\text{arg}\gamma)    
    \right],&
\end{flalign*}
while the equations for $d\phi_k^\beta/dt$ and $d\zeta_k^\beta/dt$ read as the expressions
above after interchanging $\alpha$ and $\beta$.

It is straightforward to obtain the equation of motion for the total area $A=\sum_k A_k$ using equation \eqref{eom:Ak}. Taking into account that $x^\nu_{ik}$ are symmetric and they multiply terms antisymmetric in $(ik)$ we get
\begin{equation}
    \frac{dA}{dt} = -\mathcal{N}|\gamma| \sum_{k,i}\sum_{\sa, \sb}
     \sa\sb
    \, \linknodo{y}[ki][\alpha]\linknodo{y}[ki][\beta]\sin{\left(\frac{\linknodo{\xi}[k][\sa \sb] + \linknodo{\xi}[i][(-\sa)(-\sb)]}{2}+\text{arg}\gamma\right)}.\label{Derivada general area}
\end{equation}
\end{widetext}

This equation has some important consequences. The first is that the time evolution
of the area explicitly depends on the twist angles at the links of the
graph, showing that the twist angles also play a relevant role in the
evolution of the intrinsic geometry of space.
Also, this equation allows us to show that the time
evolution of the area is bounded, as we show below.
In particular, observe that if $\gamma=0$
 the total area of the polyhedra is a constant of motion, independently of the value
of the coupling constant $\lambda$. 

Indeed, given the above bounds \eqref{eq:bounds_xy} for $y_{ij}$, the absolute
value of \eqref{Derivada general area} yields
\begin{equation}
  \left|\frac{1}{\mathcal{N}}\frac{dA}{dt}\right| \leq \sum_{i,j}\sum_{\sa,\sb} 4|\gamma|A_i A_j
  = 16|\gamma|A^2\label{inequality},
\end{equation}
after using the triangle inequality. As a consequence, 
the bounds of the total area of the polyhedron do not depend
on the coupling constant $\lambda$. This makes sense, since $\lambda$
is the factor that multiplies the terms containing
$E_{ij}^{\alpha}E_{ij}^{\beta}$, that Poisson commutes with the total area
\cite{returnSpinor,2vertexN}.

Performing a
change of variables $d\tau := \mathcal{N}(t)dt$, and considering the
limiting cases of the inequality \eqref{inequality}, it follows, first,
  that if $A(\tau)$ vanishes at some point, then it vanishes everywhere. This
  is consistent with the positivity of $A$ in the phase space,
  since if the initial total area $A$ of a polyhedron is
positive, then it is bounded from below so that $A(\tau)$ is positive at finite $\tau$.
Moreover,
given an initial area $A(0)= A_0 > 0$,
then $A(\tau)$, which is thus positive,
  satisfies the inequalities
\begin{equation}
  \frac{1}{A_0} - 16|\gamma||\tau| \leq
\frac{1}{A(\tau)} \leq \frac{1}{A_0} + 16|\gamma||\tau|.\label{boundaries AREA}
\end{equation}
As a result we have that
  \begin{equation}\label{eq:below}
\frac{A_0}{1 +16A_{0}|\gamma||\tau|} \leq A(\tau),  
\end{equation}
and that there is a critical
time $\tau_{\star} = \left(16A_0|\gamma|\right)^{-1}$ such that
\begin{equation} |\tau| < \tau_\star \implies \frac{A_0}{1 +
16A_{0}|\gamma||\tau|} \leq A(\tau) \leq \frac{A_0}{1 -
16A_{0}|\gamma||\tau|}.\label{eq:above}
\end{equation}

The inequality \eqref{eq:below} provides a minimal value of the total area
  for finite values of $\tau$. These bounds improve the global bound at
  zero ($A(\tau)>0$).
On the other hand, from \eqref{eq:above},
$A$ is also bounded from above
in the time interval $(-\tau_{\star},\tau_\star)$, and therefore the possible divergencies (known to exist from previous works \cite{Eneko}) will
occur out of that region of time (see figure \ref{figure_graficas}).
Observe that the $\gamma=0$ case is recovered in \eqref{eq:above},
yielding a constant $A(\tau)=A_0$ for all $\tau$, since $\tau_\star\to \infty$.

One can proceed in a similar manner to infer upper and lower bounds
for each of the functions $A_k(\tau)$.
From the absolute value of \eqref{eom:Ak} we obtain
  \begin{align*}
    \left|\frac{dA_k}{d\tau}\right|
    \leq &\sum_{i,\sa,\sb} |\lambda| 4 A_i A_k+ \sum_{i,\sa,\sb}|\gamma| 4 A_i A_k\\
    &=16 A A_k (|\lambda| + |\gamma|).
  \end{align*}
  As before, from this relation it follows that if $A_k(\tau)$ for some $k$ vanishes at some
  point, then $A_k$ vanishes everywhere. Thence, taking $A_k(\tau)>0$ for all $k$
  is consistent with the above construction.
  Moreover, for $|\tau|\leq \tau_\star$, using the second inequality in \eqref{eq:above} we obtain
  \[
    |\tau|\leq \tau_\star\implies \left|\frac{d\ln A_k}{d\tau}\right|  \leq  \frac{16 A_0 (|\lambda|+|\gamma|)}{1-16 A_0|\gamma||\tau|}.
  \]
  Therefore, if $\gamma\neq 0$, it holds
\begin{align} |\tau| < \tau_\star \implies
 &A_{k}(0)(1- 16A_0
   |\gamma| |\tau|)^{\left(1 + \frac{|\lambda|}{|\gamma|}\right)}  
   \label{bound:Ak}\\
              & \leq A_{k}(\tau)\leq A_{k}(0)(1- 16A_0 |\gamma||\tau|)^{-\left(1 +
\frac{|\lambda|}{|\gamma|}\right)},\nonumber
\end{align}
while for the case $\gamma = 0$ we have, for  all $\tau$,
\begin{equation}\label{bound:Ak_g}
  \gamma = 0 \implies A_{k}(0)e^{-16 A_0 |\lambda| |\tau|} \leq A_{k}(\tau) \leq A_{k}(0)e^{16 A_0 |\lambda| |\tau|}.
\end{equation}
Thus, in the case $\gamma=0$ the area of the individual faces is
bounded from below by a positive quantity at any finite time
(it is locally bounded from below by positive numbers), improving the initial bound
  found above.

These equations show the relevance of the coupling
constant $\lambda$ when separately studying the different faces of the
polyhedra. This is something expected since $\lambda$ modulates the
interchange of area between the different faces. 

At this point it is worth checking if the sum over all the faces of \eqref{bound:Ak}
  and \eqref{bound:Ak_g} improve, or not, the bounds for the total area $A$ found
  in \eqref{eq:above}. The short answer is no, that is, the inequalities in  \eqref{eq:above}
  imply the sum of the inequalities in  \eqref{bound:Ak}
  and \eqref{bound:Ak_g} over the faces. To prove that, we start with
  the case $\gamma=0$. The sum of \eqref{bound:Ak_g}
  provides
  \[
    A_0 e^{-16 A_0 |\lambda| |\tau|} \leq A(\tau) \leq A_0 e^{16 A_0 |\lambda| |\tau|},
  \]
  which is clearly superseded by $A(\tau)=A_0$. For $\gamma\neq 0$, the sum over the faces
  of \eqref{bound:Ak} yields
  \begin{align*} |\tau| < \tau_\star \implies
    &A_0(1- 16A_0|\gamma| |\tau|)^{\left(1 + \frac{|\lambda|}{|\gamma|}\right)}  
      \nonumber\\
    & \leq A(\tau)\leq A_0(1- 16A_0 |\gamma||\tau|)^{-\left(1 +
      \frac{|\lambda|}{|\gamma|}\right)}.\nonumber
  \end{align*}
  In this case it is straightforward to check that the two inequalities are
  implied by those in \eqref{eq:above} by noticing that $\Upsilon:=16 A_0|\gamma||\tau|$
  satisfies $0\leq \Upsilon <1$ for $|\tau| < \tau_\star$, and thus
  $(1-\Upsilon)\leq 1$, and also $(1-\Upsilon)^{1+\Omega}\leq (1-\Upsilon)$
  for any $\Omega\geq 0$, so that we have
  \[
    \frac{1}{1-\Upsilon}\leq (1-\Upsilon)^{-(1+\Omega)}.
    \]
  On the other hand, because of  $0<1-\Upsilon^2 = (1-\Upsilon)(1+\Upsilon)\leq 1$, we also have
  $1-\Upsilon\leq (1+\Upsilon)^{-1}$, and therefore
    \[
  (1-\Upsilon)^{1+\Omega}\leq \frac{1}{1+\Upsilon}.
    \]
  
Notice that the presence of a lower bound for the total area at finite times agrees with the analytical results obtained in the $\U(N)$ symmetry reduced sector of the model \cite{livinebenito,returnSpinor,2vertexN,FrameBasis,Alvaro,garay_two-vertex_2025}, where the cosmological behavior of the dynamics of the two-vertex model was studied, as well as with the numerical analysis carried out in \cite{Eneko} for the general case.

\begin{figure}
    \centering
    \includegraphics[width=0.48\textwidth]{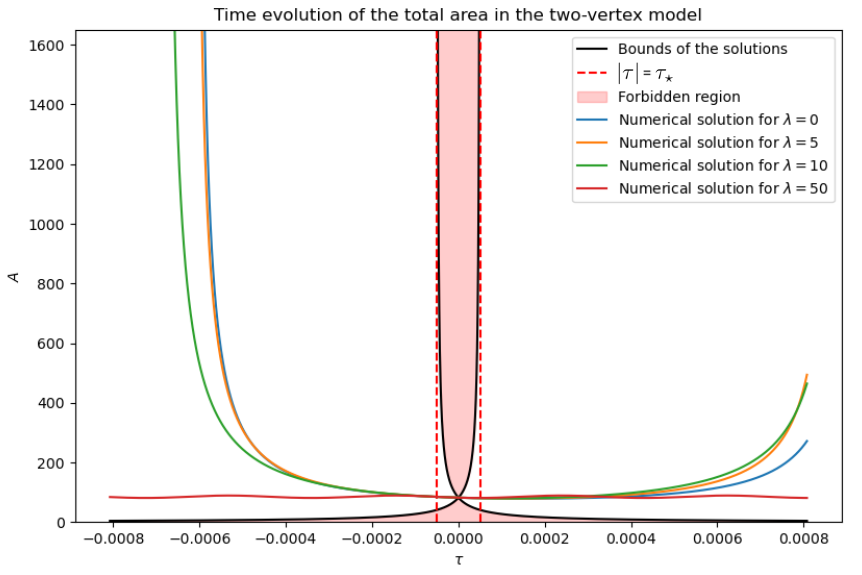}
    \centering
    \caption{Numerical solutions of $A(\tau)$. Given the bounds found, $A(\tau)$ never crosses the forbidden region. The colored curves correspond to numerical solutions with the same random initial condition for the spinors (satisfying closure and matching constraints) with $\gamma = 15$ and varying $\lambda$.}
    \label{figure_graficas}
\end{figure}

\section{Gauge fixing with the $\zeta$-variables}
\label{Sec:gauge}

In the previous section we have used the
$\zeta$-variables for the study of the dynamics in the two vertex model,
deducing some general results that are harder to establish
when directly working with spinors. In this section, we are going back to general graphs to show the relevance of these variables in the process of gauge fixing, extending the results of \cite{garay_two-vertex_2025} for  the 4-valent two-vertex model to any graph.

In \cite{garay_two-vertex_2025}, a new set of variables were introduced
to perform the gauge fixing in the two vertex model.
That construction turned out to be very useful in order to perform different reductions of the model and identify them with anisotropic and inhomogeneous sectors. However, that prescription was restricted to a two-vertex graph
with four links.
Using  the $\zeta$-variables we can generalize that construction
to an arbitrary graph.

The aim is to fix completely the gauge in a convenient way and work with the minimum number of canonical variables in phase space.

In a general graph with $N$ links and $n$ nodes, there exist $N$ matching constraints and $n$ closure
constraints. Each matching constraint is a real scalar equation and
each closure constraint is a real vector equation. That makes a total of $N+3n$ real constraints,
and 
the Hamiltonian should Poisson commute with
all of them. 
Therefore, since they commute  with each other, they are first class, and  they thus generate $N+3n$ gauge freedoms.
Taking into account that the initial phase space has dimension $8N$, that correspond to the original $2N$ spinors, the dimension of the phase space covering the physical degrees of freedom is thus $8N - (2N + 6n) = 6(N-n)$\, (note that the $6N$ degrees of freedom correspond to the reduced $\zeta$-variables).

In \cite{garay_two-vertex_2025}, the gauge fixing procedure
was studied for $n = 2$ and $N=4$, where the physical degrees of
freedom (after gauge fixing) were encoded in a set of canonical variables.
We will now generalize that reduction to any
value of $n$ and $N$, and we will show that, indeed, the new variables  provide a generalization of those introduced in \cite{garay_two-vertex_2025}.

\subsection{Matching constraint}

The initial $8N$ degrees of freedom in a graph
are represented by the variables \eqref{variables a priori} at each of
the $N$ links of the graph. As already discussed in Section \ref{sec:variables}
using the matching constraint we reduce the number of variables
to the six $\U(1)$ gauge invariant quantities on each link $i$
given by the reduced $\zeta$-variables \eqref{variables matching}, where in order to be consistent with the choices made in  
 \cite{garay_two-vertex_2025} we shall fix
the gauge relative to the matching constraint to
\begin{equation}
       \varepsilon_i^{s} - \varepsilon_i^{t} = 0.    
\end{equation}

So far, this procedure
  is clearly  valid for any value of $N$, and we have reduced the number of variables
to $6N$. 

\subsection{Closure constraint}
The aim now is 
to reduce the remaining $6n$ spurious degrees of
freedom resulting from the $3n$ closure constraints of the graph and the $3n$ associated gauge freedoms. To do that we follow
a geometrical procedure analogous to the one performed in
\cite{garay_two-vertex_2025}.

We already know that the closure constraints at the nodes of the graph
translate into having well defined convex polyhedra placed at
them. Therefore, we can assume that fact (so that we reduce $3n$
additional degrees of freedom) and then fix the corresponding gauge
freedom, which is associated with the transformations
$\ket{z^{\nu}} \to h^\nu \ket{z^{\nu}}$, with $h^\nu\in  \SU(2)$, for each node $\nu$ of the graph.
Such transformations translate into
global rotations of the polyhedron, so that fixing the gauge freedom is
equivalent to establishing a way to fix the orientation of each
polyhedron in the graph.

Therefore, since we have already defined $2N$
independent variables $\left\{A_i,\Phi_i\right\}_{i=1}^{N}$, the
remaining $4N-6n$ physical degrees of freedom will be encoded by some
variables that, once known, should allow us to reconstruct the
polyhedra and their orientation according to the gauge fixing we decide to establish.

This process is best illustrated by the $n = 2$ and $N=4$ (two-vertex model with four links) case as
worked out in \cite{garay_two-vertex_2025}, which we summarize in the following.

To fix the orientation of a tetrahedron, one starts by taking two normal
vectors ($\vec{X}_1$ and $\vec{X_2}$) and rotating the polyhedron
until $\vec{X}_1 + \vec{X}_2$ is parallel to the $z$-axis and pointing
towards positive  $z$. Alternatively we \emph{fix}, at this node, the $z$-axis
to point towards  $\vec{X}_1 + \vec{X}_2$. In virtue of the closure constraint,
this implies that $\vec{X}_3 + \vec{X}_4$ will also be parallel to the
$z$-axis and pointing in the opposite direction.
The remaining freedom consists of
rotations around the $z$-axis. Thus, we have to complete the
gauge fixing with an additional condition. In \cite{garay_two-vertex_2025} it
is imposed that the projections of $\vec{X}_1$ and $\vec{X}_3$ on the
$XY$-plane must subtend opposite angles relative to the
$x$-axis. Since there are still two possible orientations fitting the
two conditions, it is additionally required the second component of
$\vec{X}_1$ to be positive. This can be stated in terms of the
$\zeta$-variables in the form
\begin{equation}
  \phi_{1} + \phi_{3} = 0, \hspace{5mm} \phi_1 \geq 0. 
  \label{gauge fix rot}
\end{equation}

This description is applied to the two  tetrahedra of the case $n =2$, $N=4$, completely fixing their orientation. It remains to
define the $4N-6n = 4$ variables that, together with
$\left\{A_i,\Phi_i\right\}_{i=1}^{4}$, contain all the information needed to
reconstruct the polyhedra given this orientation fixing.

In \cite{garay_two-vertex_2025},  the two nodes were denoted $\alpha$ and $\beta$, so that these four remaining variables were denoted
$x^{\alpha},\varphi^{\alpha}$ and $x^{\beta},\varphi^{\beta}$.
The variables $x^\nu$ for $\nu\in\{\alpha,\beta\}$ were defined as
\begin{align*}
  &x^\nu = |\vec{X}_1^\nu + \vec{X}_2^\nu|,
   \end{align*}
so that they correspond to
the projections of $\vec{X}^\nu_1 + \vec{X}^\nu_2$ on the $z$-axis according
to the gauge fixing just established,
while $\varphi^\nu$ were defined to be
the angles between the projections of $\vec{X}^\nu_1$ and $\vec{X}^\nu_3$ on
the corresponding $XY$-planes.

The variables $x^\nu$ can be written in terms of the $\zeta$-variables by
\begin{equation}
    x^\nu= \zeta_{1}^\nu + \zeta_{2}^\nu = -\zeta_{3}^\nu - \zeta_{4}^\nu,
\end{equation}
where the second equality follows from the closure constraint,
while
\begin{equation}
    \varphi^\nu= \phi_1^\nu - \phi^\nu_{3},\label{eq:varphi}
\end{equation}
according to the gauge fixing.
Besides, the variables $\{x^\nu,\varphi^\nu\}$ satisfy,
\begin{equation}
    \left\{x^{\nu_1},\varphi^{\nu_2}\right\} = \delta^{\nu_1 \nu_2},
\end{equation}
where $\nu_1,\nu_2\in\left\{\alpha,\beta\right\}$. 

The $12$ variables 
\begin{equation}
     \left\{A_{i},\Phi_{i}\right\}_{i=1}^{4} \cup \left\{x^\alpha,x^\beta,\varphi^\alpha,\varphi^\beta\right\},
\end{equation}
are already independent and thus valid to describe the physical
degrees of freedom of the theory.
Now, since we are interested in
generalizing these variables to any graph, it is useful to
understand how they allow us to reconstruct the tetrahedra
once we have fixed the gauge.

We reason as follows. Fix $\nu$ to be $\alpha$ or $\beta$.
  Once we are given $\varphi^\nu$, using equations
\eqref{eq:varphi} and \eqref{gauge fix rot} (for the specific node)
we directly obtain $\phi^\nu_1$ and
$\phi^\nu_3$. On the other hand, given $x^\nu =
\zeta^\nu_1 + \zeta^\nu_{2}$ together with $A_1$ and $A_2$,
we determine the angle between the vectors $\vec{X}^\nu_1$ and
$\vec{X}^\nu_2$
using the fact that $\vec{X}^\nu_1+\vec{X}^\nu_2$ is parallel to the
$z$-axis pointing towards positive $z$. This, together with
$\phi_1^\nu$, provides all the information needed to recover $\vec{X}^\nu_1$
and $\vec{X}^\nu_2$. Applying the same reasoning, knowing $x^\nu = -
\zeta^\nu_3 - \zeta^\nu_4$, together with $A_3$ and $A_4$,
allows us to recover the angle between the vectors $\vec{X}^\nu_3$ and
$\vec{X}^\nu_4$. Then, we only need $\phi^\nu_{3}$ to recover $\vec{X}^\nu_3$.
Finally, $\vec{X}^\nu_4$ is obtained  through the closure constraint.
It only remains to follow the same procedure for the other node.

\subsection{Generalization to arbitrary graphs}

It turns out that we can use the $\zeta$-variables to generalize the
variables $\{x^\alpha,x^\beta,\varphi^{\alpha},\varphi^ \beta\}$
 to general graphs, with
more than two nodes and polyhedra with more than four faces. At first glance, this may seem straightforward, since one
could again fix the orientation of polyhedra by taking the sum
$\vec{X}_1 + \vec{X}_2$, forcing it to be parallel to the $z$-axis,
and then imposing $\phi_1 + \phi_3 = 0$ and $\phi_1 \geq 0$, just like
in the preceding section. However, this only makes sense, of course,
when, first, 
$\vec{X}_1 + \vec{X}_2 \neq 0$,
and second, neither  $\vec{X}_1$ nor $\vec{X}_3$ are
  parallel to the $z$-axis.

When all the polyhedra are tetrahedra, both conditions are always satisfied
regardless of the choice of labelling of the vectors $\vec X_i$.
The first, $\vec{X}_1 + \vec{X}_2 \neq 0$, holds because there are no
parallel faces in
a three-dimensional tetrahedron.
Then, once
$\vec{X}_1 + \vec{X}_2$ is oriented
along the $z$-axis, since we
assume $\vec{X}_i \neq 0$,
neither $\vec{X}_1$ nor
$\vec{X}_2$ are parallel to the $z$-axis.
Therefore, $\vec{X}_1$ satisfies the second condition.
Following a similar reasoning, the closure constraint
implies that $\vec{X}_3$ cannot be parallel to the $z$-axis.

When we allow polyhedra to have more than four faces, however, these two conditions may
in principle be violated
depending on the numbering chosen for the vectors. Nevertheless, it is always
possible to label the 
vectors $\vec{X}_i$ in a way such that the two conditions are met 
simultaneously. First, since
we can always find two faces of the polyhedron
that are not parallel,
we take that pair and label them  by $1$ and $2$,
so that the first condition holds.
Again,
after orienting $\vec{X}_1 +\vec{X}_2$ along the $z$-axis,
$\vec{X}_1$ will not be parallel to it (see figure \ref{figure_fixing1}).
Now, there exists a third vector not parallel to the  $z$-axis since
otherwise the faces different from $1$ and
$2$ would be all parallel and hence no polyhedron would
exist. If we denote the valence of the node $\nu$ by $N_{\nu}$, it suffices to
label that vector as\footnote{Let us note that
  in \cite{garay_two-vertex_2025} (for $N=4$) the third vector is labelled as $\vec{X}_3$
  instead of $\vec{X}_4$, but that only accounts for a trivial relabelling
    with no further consequences.} 
$\vec{X}_{N_{\nu}}$ to ensure that the two 
conditions hold.

\begin{figure}
    \centering
    \includegraphics[width=0.47\textwidth]{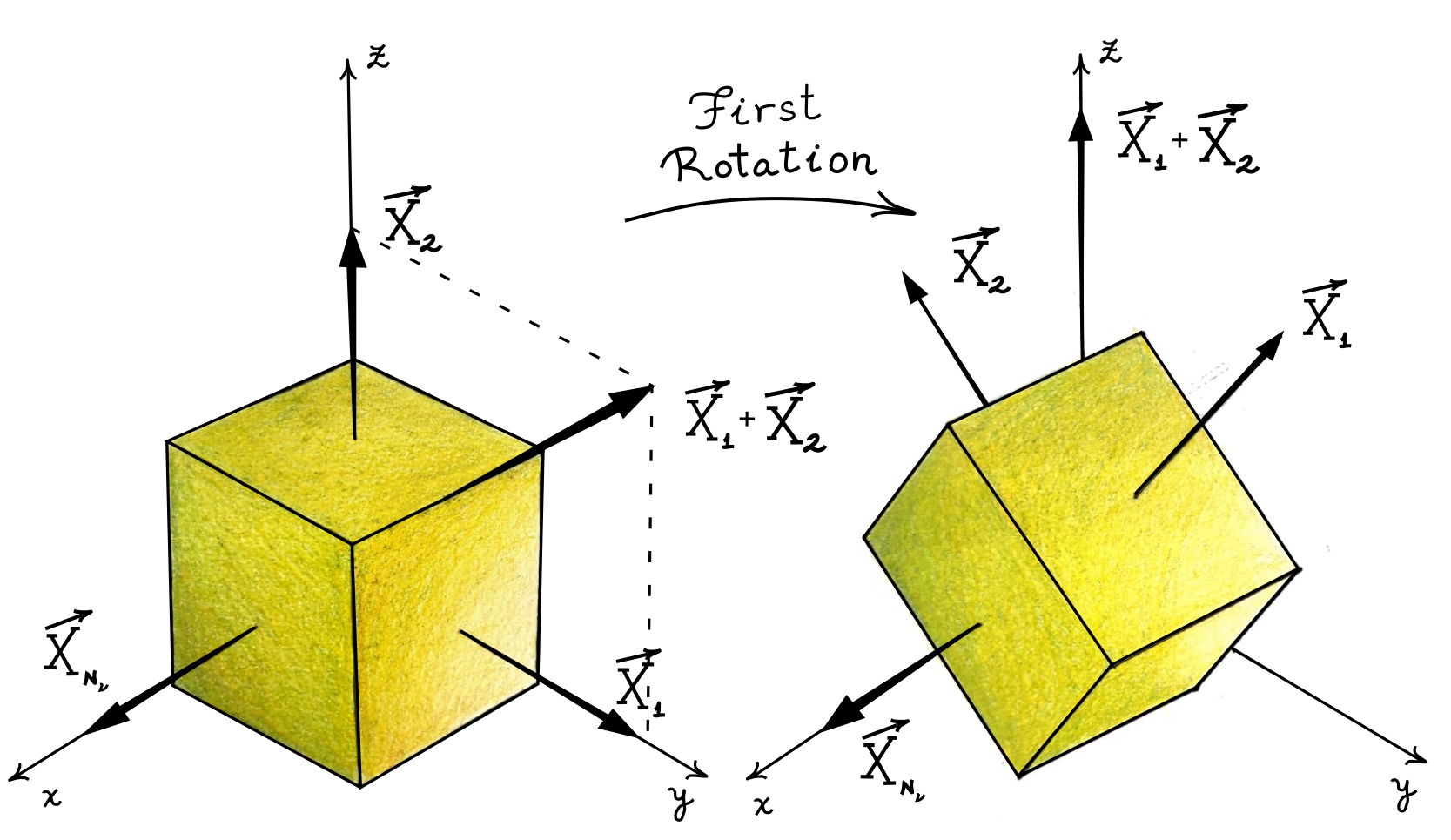}
    \centering
    \caption{Representation of the first step in the gauge fixation. The polyhedron is rotated until $\vec{X}_1 + \vec{X}_2$ is oriented along the $z$-axis, pointing towards the positive direction.}
    \label{figure_fixing1}
\end{figure}

\begin{figure}
    \centering
    \includegraphics[width=0.47\textwidth]{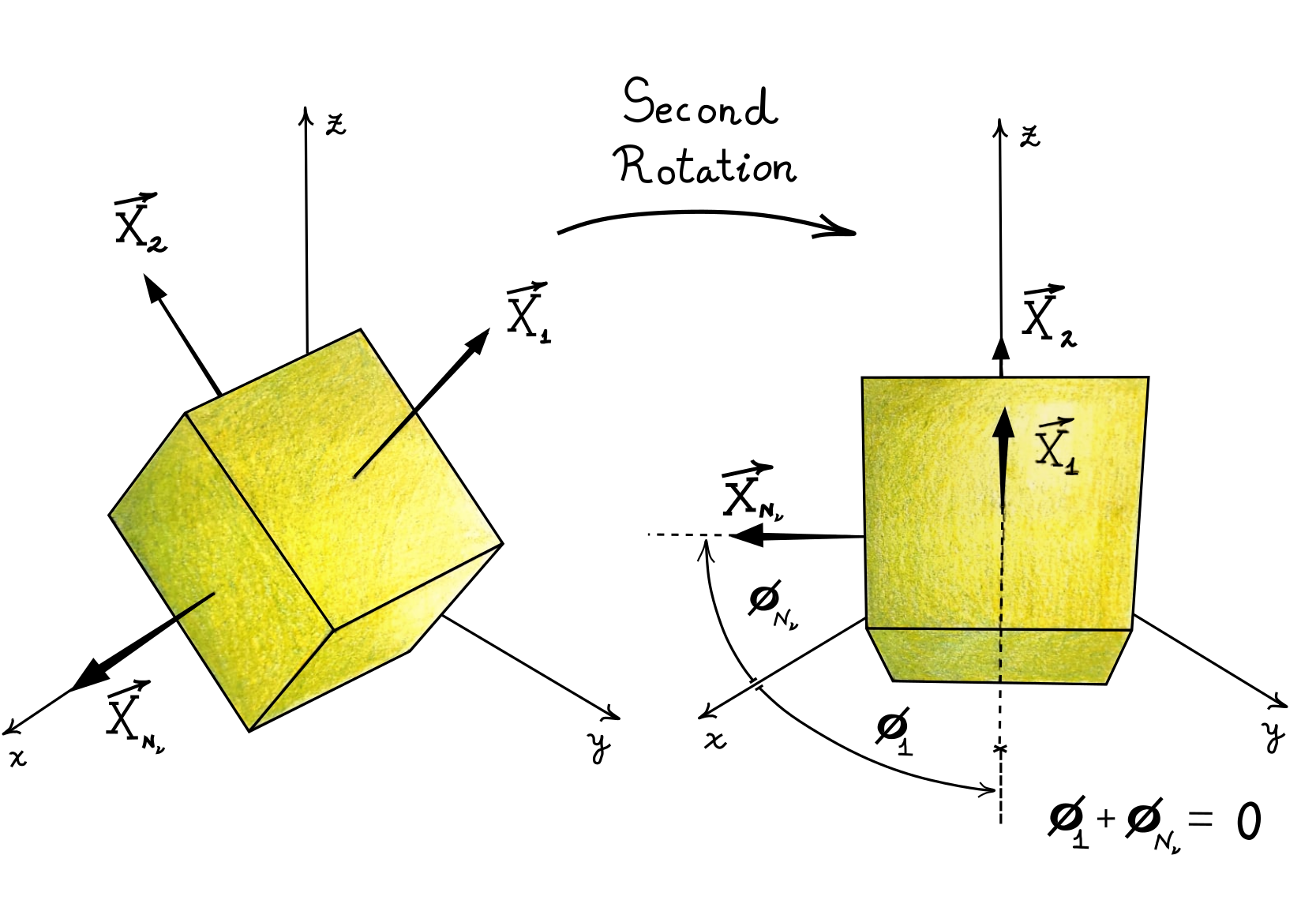}
    \centering
    \caption{Representation of the second step in the gauge fixation we described. Once the polyhedron has been oriented in a way such that $\vec{X}_1 + \vec{X}_2$ points towards the positive direction of the $z$-axis, the polyhedron is rotated around it until the projections of $\vec{X}_1$ and $\vec{X}_{N_\nu}$ on the $XY$-plane are such that $\phi_1 +\phi_{N_\nu} = 0$ and $\phi_1 \geq 0$.}
    \label{figure_fixing2}
\end{figure}

Now, we have to find the generalization of the variables
$\{x^\nu,\varphi^{\nu}\}$
to these cases. First, we use the vector labelled by $\vec{X}_{N_{\nu}}$ to fix the
rotational gauge through the condition 
\begin{equation} \phi_{1}^{\nu} + \phi_{N_{\nu}}^{\nu} = 0, \hspace{5mm}
  \phi^{\nu}_{N_{\nu}} \geq 0\label{gauge rota N}
\end{equation}
at each node $\nnu$ of the graph (see figure \ref{figure_fixing2}).
Following a reasoning analogous to that at the end of
the previous section we can proceed to the definition of
the $4N-6n$ remaining variables as follows:

\begin{enumerate}
\item First, define the variables
  $x_1^{\nnu} := \zeta_1^{\nnu} + \zeta_2^{\nnu}$ and
  $\varphi_1^{\nnu} := \phi_1^{\nnu} - \phi_{N_{\nu}}^{\nnu}$ at each node $\nu$.
  Given the values of $A_i$, i.e. the areas of the faces,
  from $x_1^{\nnu}$ and $\varphi_1^{\nnu}$ we determine
$\vec{X}_1^{\nnu}$, $\vec{X}_2^{\nnu}$, $\phi_{N_{\nu}}^{\nnu}$.
\item Next, define
  $x_2^{\nnu} :=\sum_{j=3}^{{N_{\nu}}-1}\zeta_{j}^{\nnu}$ and
  $\varphi_2^{\nnu} :=\phi_{{N_{\nu}}-1}^{\nnu} - \phi_{N_{\nu}}^{\nnu}$.
  Since we already know
$\phi_{N_{\nu}}^{\nnu}$ from the previous step, from $\varphi_2^\nu$ we obtain
$\phi_{{N_{\nu}}-1}^{\nnu}$. Additionally, in virtue of the closure
constraint, from $x_1^{\nnu}$ and $x_2^{\nnu}$ we recover
$\zeta_{N_{\nu}}^{\nnu}$. Hence, given $\phi_{N_{\nu}}^{\nnu}$, $\zeta_{N_{\nu}}^{\nnu}$ and
$A_{N_{\nu}}^{\nnu}$ we determine $\vec{X}_{N_{\nu}}^{\nnu}$. Therefore, at this step
we have already defined the quantities needed to recover
$\vec{X}_1^{\nnu}$,
$\vec{X}_2^{\nnu}$, $\vec{X}_{N_{\nu}}^{\nnu}$, $\phi_{{N_{\nu}}-1}^{\nnu}$.
\item Now, given $x_3^{\nnu} := \sum_{j=3}^{{N_{\nu}}-2} \zeta_{j}^{\nnu}$ and
  $\varphi_3^{\nnu} := \phi_{{N_{\nu}}-2}^{\nnu} - \phi_{{N_{\nu}}-1}^{\nnu}$, we recover $\phi_{{N_{\nu}}-2}^{\nnu}$
  and $\vec{X}_{{N_{\nu}}-1}^{\nnu}$ in virtue, again, of the closure constraint.
  Thus, at this step we have defined the variables needed to recover
  $\vec{X}_1^{\nnu}$, $\vec{X}_2^{\nnu}$, $\vec{X}_{N_{\nu}}^{\nnu}$, $\vec{X}_{{N_{\nu}}-1}^{\nnu}$, $\phi_{{N_{\nu}}-2}^{\nnu}$.
\item  Iterate step 3
  until we have defined the $2N_\nu-6$ variables encoding the physical degrees of
  freedom at node $\nu$ that remain after the matching constraint reduction.
\end{enumerate}
To sum up, at the end of the iteration we end up having defined
the variables
\begin{align}
  &x_1^{\nnu} := \zeta_1^{\nnu} + \zeta_2^{\nnu},  \quad \varphi_1^{\nnu} := \phi_1^{\nnu} - \phi_{N_{\nu}}^{\nnu},\nonumber\\
  &x_{i}^{\nnu} := \sum_{j=3}^{{N_{\nu}}+1-i}\zeta_{j}^{\nnu}, \qquad  i= 2,\ldots,{N_{\nu}}-3, \label{Generalized_DV1}\\
  &\varphi_{i}^{\nnu} := \phi_{{N_{\nu}}+1-i}^{\nnu} - \phi_{{N_{\nu}}+2-i}^{\nnu}, \qquad i= 2,\ldots,{N_{\nu}}-3.\nonumber
\end{align}

These variables can be proven to be independent and canonical, since they satisfy the Poisson brackets
\begin{equation}
    \left\{x_i^{\nu_1},\varphi_{j}^{\nu_2}\right\} = \delta_{ij}\delta^{\nu_1 \nu_2}.
\end{equation}
Note, as consistency check, that this yields $2N_\nnu-6$ variables at each node, so the total number of 
$(x_{i}^{\nnu}, \varphi_{i}^{\nnu})$ variables is
$$
\sum_{\nnu=1}^n(2N_\nnu-6)=4N-6n,
$$
where $\sum_{\nnu=1}^nN_\nnu=2N$ is assumed (all the links connect two different nodes).
This generalization allows us to work with
the minimum number of variables we need in order to describe all the 
physical information of the classical state of the
polyhedra dual to the graph. 

\section{Conclusions}

In this work we have used a  canonical parametrization of twisted geometries
to construct a
 framework that provides some  conceptual and technical advantages in the study of LQG on a fixed graph. In particular, the reduced $\zeta$-variables allow for a convenient description of the dynamics in the two-vertex model and offer a natural setting to implement gauge fixing in general graphs, extending previous constructions restricted to the four-link case of the two-vertex model \cite{garay_two-vertex_2025}.

One of the main results of this paper is the derivation of analytical bounds for the evolution of the total area in the two-vertex model. In particular, inequality \eqref{eq:below} provides a non-trivial lower bound for the total area at finite values of the evolution parameter. This result improves the 
positivity bound and shows that, starting from an initial configuration, the total area remains strictly positive along the evolution. Such behavior is suggestive of a ``bounce''-like scenario, reminiscent of results found in loop quantum cosmology \cite{Bojowald:2001xe,Ashtekar:2006rx,Ashtekar:2006wn}, where singularities are replaced by non-vanishing minimal volumes. While our analysis is purely classical and restricted to the two-vertex model, the existence of these bounds points towards a deeper connection that deserves further investigation.

It is worth emphasizing that these results were not previously accessible through analytical methods. Indeed, the existence of such bounds had only been observed in numerical studies in the general case \cite{Eneko}, as well as in symmetry-reduced sectors such as those with U(N) invariance \cite{returnSpinor,FrameBasis,Alvaro,2vertexN,livinebenito}. The $\zeta$-variables provide 
an appropriate framework to derive these features analytically and to understand their origin in terms of the underlying geometric degrees of freedom. Besides, the generalization of the parametrization given in \cite{garay_two-vertex_2025} to an arbitrary number of links
 opens the door to systematically explore the emergence of
anisotropic or inhomogeneous cosmological models within this framework.

Therefore, we expect that the $\zeta$-variables will prove useful in further studies of both classical and quantum dynamics in LQG, as well as in the investigation of effective cosmological models derived from it.

 \acknowledgments
 This work is supported by the Basque Government Grant IT1628-22, and by the Grant PID2021-123226NBI00 (funded by MCIN/AEI/10.13039/501100011033 and by ``ERDF A way of making Europe''). Additionally, S. R. acknowledges financial support from MIU (Ministerio de Universidades, Spain) fellowship FPU FPU23/01491.


\bibliography{references.bib}

\end{document}